\documentclass[letterpaper,twoside,twocolumn,english,superscriptaddress,showpacs]{revtex4}
\usepackage[T1]{fontenc}
\usepackage[latin9]{inputenc}
\usepackage{babel}
\usepackage{bm}
\usepackage{amsmath}
\usepackage{amssymb}
\usepackage{graphicx}
\usepackage{esint}
\usepackage[unicode=true,pdfusetitle,
 bookmarks=false,
 breaklinks=false,pdfborder={0 0 1},backref=false,colorlinks=false]
 {hyperref}

\makeatletter

\@ifundefined{textcolor}{}
{%
 \definecolor{BLACK}{gray}{0}
 \definecolor{WHITE}{gray}{1}
 \definecolor{RED}{rgb}{1,0,0}
 \definecolor{GREEN}{rgb}{0,1,0}
 \definecolor{BLUE}{rgb}{0,0,1}
 \definecolor{CYAN}{cmyk}{1,0,0,0}
 \definecolor{MAGENTA}{cmyk}{0,1,0,0}
 \definecolor{YELLOW}{cmyk}{0,0,1,0}
 }

\makeatother

\begin{document}

\title{Topological Invariants of Metals and Related Physical Effects}

\author{Jianhui Zhou }

\affiliation{Beijing National Laboratory for Condensed Matter Physics and Institute
of Physics, Chinese Academy of Sciences, Beijing 100190, China}

\author{Hua Jiang }

\affiliation{International Center for Quantum Materials, Peking University , Beijing
100871, China}

\author{Qian Niu }

\affiliation{International Center for Quantum Materials, Peking University , Beijing
100871, China}

\affiliation{Department of Physics, The University of Texas, Austin, Texas 78712-0264,
USA }

\author{Junren Shi }

\affiliation{International Center for Quantum Materials, Peking University , Beijing
100871, China}
\begin{abstract}
The total reciprocal space magnetic flux threading through a closed
Fermi surface is a topological invariant for a three-dimensional metal.
For a Weyl metal, the invariant is non-zero for each of its Fermi
surfaces. We show that such an invariant can be related to magneto-valley-transport
effect, in which an external magnetic field can induce a valley current.
We further show that a strain field can drive an electric current,
and the effect is dictated by a second class Chern invariant. These
connections open the pathway to observe the hidden topological invariants
in metallic systems.
\end{abstract}


\maketitle
There have been great successes in classifying insulators using topological
invariants, such as the TKNN number for quantum Hall insulators\cite{Thouless1982},
and the $Z_{2}$ index for insulators with time-reversal symmetry~\cite{Kane2005}.
The classification can even be extended to the more general systems
such as superconductors~\cite{Ryu2010}. In all of these systems,
there exist gaps in their excitation spectrums, which provide topological
protection, and resulting in robust physical effects. 

For metallic systems, there are also topological invariants. An apparent
one is the total volume enclosed by Fermi surface in a non-magnetic
metal, according to Luttinger's theorem. One may find other topological
invariants, and in some cases, even relate them to observable physical
effects~\cite{volovik,Oshikawa2000}. For instance, in a two dimensional
(2D) metallic system with time reversal symmetry, an electron moving
around the Fermi cycle in the reciprocal space will acquire a total
Berry phase of either $0$ or $\pi$ per cycle. The non-trivial $\pi$
Berry phase is responsible for the unusual anti-weak localization
effect observed in compounds such as graphene or the surface states
of three-dimensional (3D) topological insulators~\cite{Lu2011}.
It can even modify the effective electron-electron interaction, giving
rise to unconventional superconducting pairing symmetries~\cite{Shi2006,Mao2011,ZhouQS}. 

For a 3D metal, an apparent topological invariant is the total reciprocal
space magnetic flux threading through a closed Fermi surface:
\begin{equation}
\Phi_{FS}\equiv\int_{FS}\bm{\Omega}\cdot\mathrm{d}\bm{S}_{F}\,,
\end{equation}
where $\bm{\Omega}\left(\boldsymbol{k}\right)$ is the Berry curvature
of the Bloch electron in band forming the Fermi surface at quasi-momentum
$\bm{k}$, defined by $\bm{\Omega}\left(\boldsymbol{k}\right)=i\left\langle \nabla_{\bm{k}}u\left(\boldsymbol{k}\right)\right|\bm{\times}\left|\nabla_{\bm{k}}u\left(\boldsymbol{k}\right)\right\rangle $,
and $\left|u\left(\boldsymbol{k}\right)\right\rangle $ is the corresponding
periodic part of Bloch wave function~\cite{Sundaram Niu,xiao chang Niu rmp}.
$\Phi_{FS}$ is quantized as 
\begin{equation}
\Phi_{FS}=2\pi n\,,\label{eq:quantized_flux}
\end{equation}
with $n$ being an integer. A nonzero quantized number $n$ indicates
the topological nature of the metal. The topological number $n$ cannot
be changed without changing the geometric topology of the Fermi surface.
It is easy to generalize the classification to the more general metallic
systems with multiple Fermi surfaces. In this case, one can assign
a topological number to each of the Fermi surfaces, and the numbers
cannot be changed without merging/splitting of the Fermi surfaces.
An example of such topologically non-trivial 3D metals is the 3D Weyl
(semi-)metal, in which the Fermi surfaces enclose Weyl nodes that
give rise to the quantized reciprocal magnetic fluxes~\cite{Nielsen Ninomiya plb,Hosur Ryu Ashvin,wan,halasz balents,burkov balents,gang xu}.
An immediate question is how the topological invariant can be related
to observable physical effects. 

In this Letter, we will show that the invariant can be related to
the magneto-valley-transport property of the 3D topological metals.
We show that a magnetic field will induce an electric current along
the direction of the magnetic field in each of the Fermi pockets.
For a 3D metal that always has an even number of Dirac points (due
to the Fermion doubling theorem), such a response will generate a
valley-current response, with the valleys corresponding to the Dirac
fermions of opposite helicities. On the other hand, using carefully
engineered strain field, one could create an effective magnetic field
that has the opposite directions in the two different valleys, generating
a real electric current response. We will show that such response
is related to a second-class Chern number defined on the Fermi surfaces
and the real space.

To see the response of the system to an applied magnetic field, we
use the semiclassical wave packet dynamics~\cite{Sundaram Niu,xiao chang Niu rmp}.
The equations of motion for the wave-packet center position $\bm{r}$
and momentum $\bm{k}$ in the presence of a magnetic field $\bm{B}$
can be written as,

\begin{align}
\dot{\boldsymbol{r}} & =\bm{v}_{n}\left(\boldsymbol{k}\right)-\dot{\boldsymbol{k}}\times\bm{\Omega}_{n}\left(\boldsymbol{k}\right)\label{eq:rdot1}\\
\hbar\dot{\boldsymbol{k}} & =-e\dot{\boldsymbol{r}}\times\boldsymbol{B}\label{eq:kdot1}
\end{align}
where $\Omega_{n}\left(\boldsymbol{k}\right)$ is the Berry curvature
of Bloch states of the $n$-th band, and $v_{n}\left(\boldsymbol{k}\right)=\partial\tilde{\epsilon}_{n}\left(\boldsymbol{k}\right)/\partial\hbar\boldsymbol{k}$
is the group velocity of Bloch electrons and $\tilde{\epsilon}_{n}\left(\boldsymbol{k}\right)$
is the band dispersion. We note that although the set of equations
is semi-classical, it is versatile in deriving results valid in quantum
regime. Actually one can establish that the result derived from the
semiclassical equations are in general valid as long as the external
fields are weak in the length scales of the lattice constants.

It is straightforward to get: 

\begin{align}
D_{n}\left(\boldsymbol{k}\right)\dot{\boldsymbol{r}} & =\bm{v}_{n}\left(\boldsymbol{k}\right)+\frac{e}{\hbar}\left(v_{n}\cdot\bm{\Omega}_{n}\right)\boldsymbol{B}
\end{align}
where $D_{n}\left(\boldsymbol{k}\right)\equiv1+(e/\hbar)\left(\boldsymbol{B}\cdot\bm{\Omega}_{n}\right)$
stands for the Berry curvature's correction to the density of states
in phase space~\cite{xiao shi Niu DOS}. We obtain a response of
the electric current $\boldsymbol{j}_{B}$ in the direction parallel
to the applied magnetic field $\boldsymbol{B}$:

\begin{align}
\boldsymbol{j}_{B} & =-e\sum_{n}\int D_{n}\left(\boldsymbol{k}\right)\dot{\boldsymbol{r}}f\left(\tilde{\epsilon}_{n}\left(\boldsymbol{k}\right)\right)\left[d\boldsymbol{k}\right]=\alpha_{B}\boldsymbol{B}
\end{align}
where the magneto-current response coefficient is defined by 

\begin{eqnarray}
\alpha_{B} & \equiv & -\frac{e^{2}}{\hbar}\sum_{n}\int\left(\bm{v}_{n}\cdot\bm{\Omega}_{n}\right)f\left(\tilde{\epsilon}_{n}\left(\boldsymbol{k}\right)\right)\left[d\boldsymbol{k}\right]\\
 & = & -\frac{e^{2}}{h^{2}}\sum_{n}\int\Phi_{\epsilon}^{(n)}f\left(\epsilon\right)\frac{d\epsilon}{2\pi}
\end{eqnarray}
where $\Phi_{\epsilon}^{(n)}\equiv\int_{\tilde{\epsilon}_{n}(\bm{k})=\epsilon}d\vec{S}\cdot\bm{\Omega}_{n}$
stands for the flux threading through the iso-energy surface at $\epsilon$
for $n$-th band, with the normal direction of the $d\vec{S}$ being
defined as the direction of the group velocity $\bm{v}_{n}(\bm{k})$,
and $f\left(\epsilon\right)$ is the Fermi distribution function,
$\left[d\boldsymbol{k}\right]\equiv\mathrm{d}\bm{k}/(2\pi)^{3}$. 

\begin{figure}

\includegraphics[width=1\columnwidth]{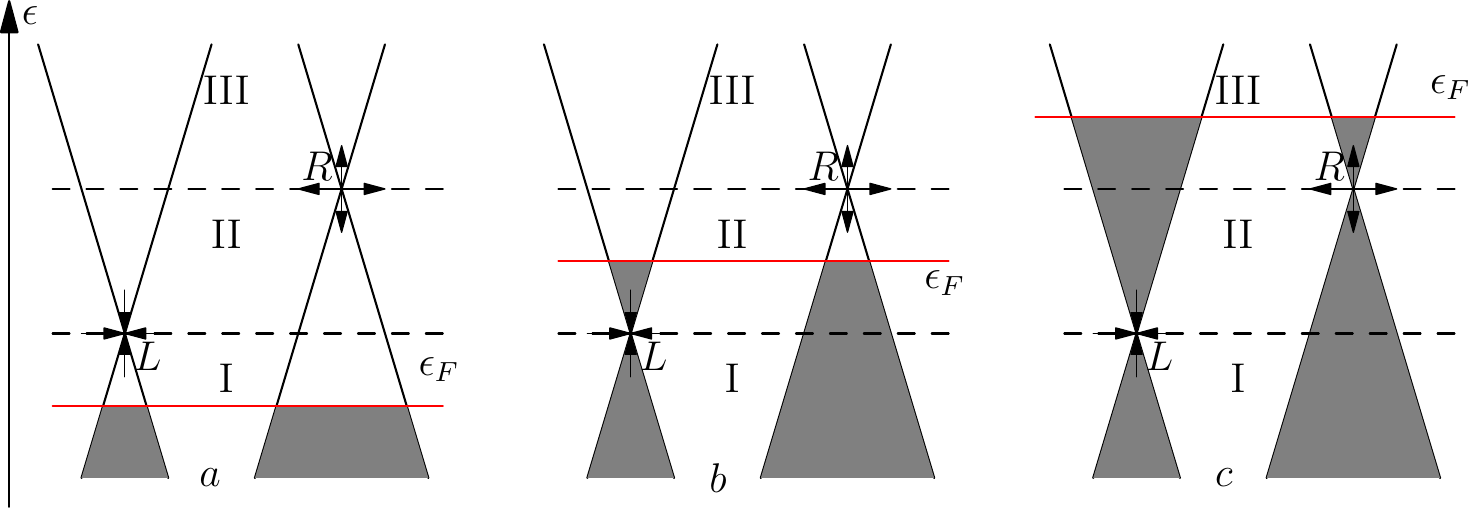}

\caption{\label{fig:flux-two-dirac}A schematic plot to show that the electric
current response to a magnetic field for a real 3D metal is always
zero because the contribution from the two Dirac points always cancel
each other. Note that for each individual Dirac point, the contributions
from the lower and the upper bands have the same sign, because the
normal directions (directions of the group velocities) of the iso-energy
surfaces are opposite.}
\end{figure}

However, for a real 3D metal, the total electric current response
is always zero. This is because the Dirac points, which are the source
of the reciprocal space magnetic field fluxes, always appear in pairs
in the reciprocal space, and in the opposite helicities, according
to the fermion doubling theorem~\cite{Nielsen1981}. As shown in
Fig.~\ref{fig:flux-two-dirac}, the reciprocal space magnetic fluxes
from such a pair of Dirac points always cancel each other at all the
energies. As a result, the magneto-current response coefficient $\alpha_{B}$
is always zero for a real 3D metal.

On the other hand, one can construct a non-zero response of valley
current for systems with nontrivial topological numbers of Fermi surfaces.
To be specific, we consider a 3D Weyl (semi-)metal~\cite{Nielsen Ninomiya plb,Hosur Ryu Ashvin,wan,halasz balents,burkov balents,gang xu}.
The simplest form of such a metal comprises of a pair of Weyl nodes
that have the reciprocal magnetic monopole charge $\pm2\pi n$. When
the two Weyl nodes are well separated in the momentum space and the
Fermi level is close to the energy of the Weyl nodes, the electrons
in the two Fermi pockets centering around the Weyl nodes can be considered
as two independent fermion species, and we can introduce a new {}``valley''
degree of freedom to label them. An external magnetic field will induced
the opposite electric currents in the two valleys at the zero temperature:
\begin{equation}
\bm{j}_{\pm}=\pm n\frac{e^{2}}{h^{2}}\left(\epsilon_{F}-\epsilon_{0}\right)\bm{B}\,,\label{eq:jpm}
\end{equation}
where $\pm$ denotes the valley index, $\epsilon_{F}$ is the Fermi
energy and $\epsilon_{0}$ is the energy of the Weyl nodes. The corresponding
valley current is:
\begin{eqnarray}
\bm{j}_{v} & = & \bm{j}_{+}-\bm{j}_{-}\,,\\
 & = & n\frac{2e^{2}}{h^{2}}\left(\epsilon_{F}-\epsilon_{0}\right)\bm{B}\,.
\end{eqnarray}

\begin{figure}
\includegraphics[width=1\columnwidth]{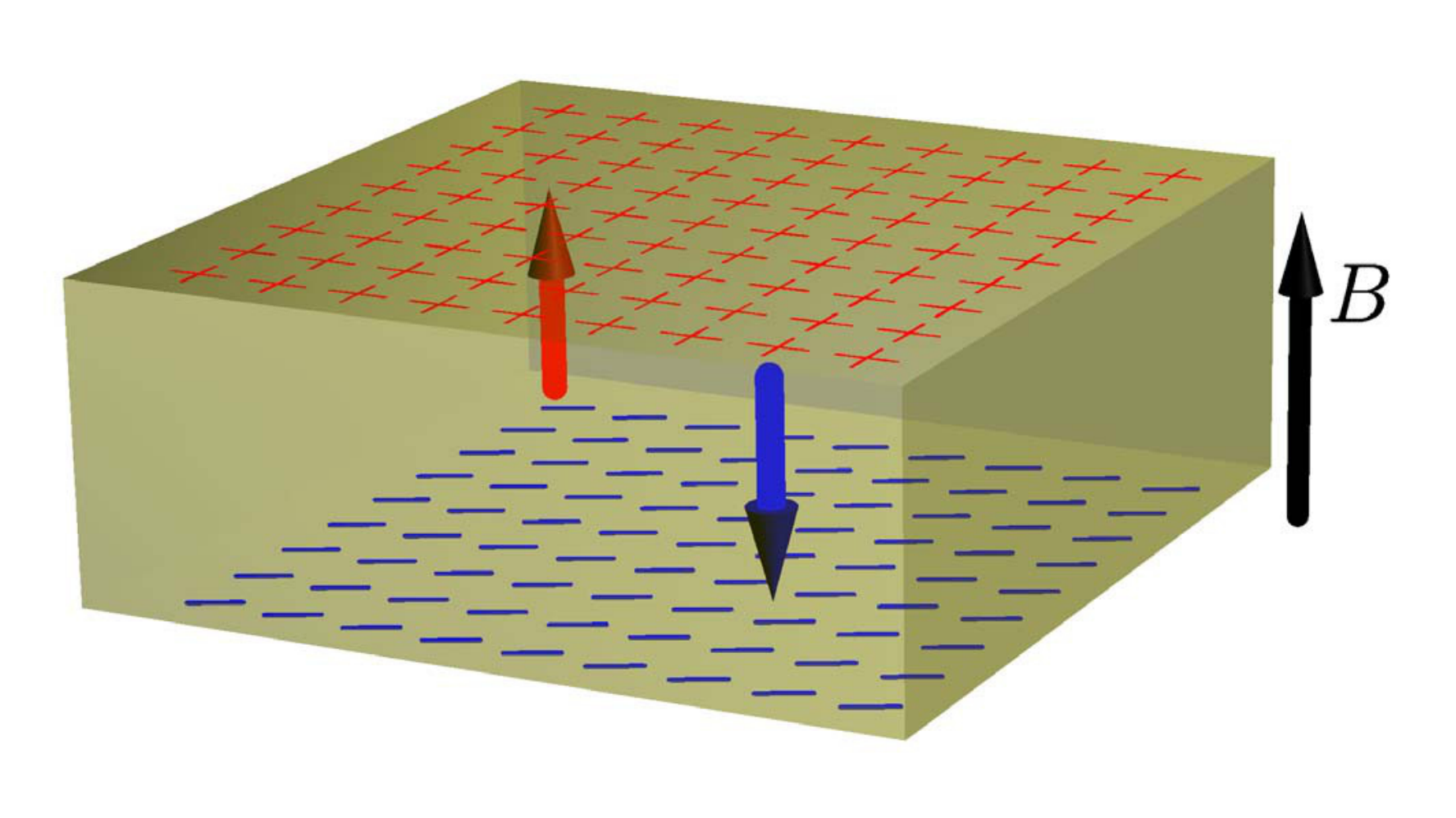}

\caption{\label{fig:valley-polarization}The magnetic field induced valley
polarization. The magnetic field $\boldsymbol{B}$ (black arrow) will
induce counter-propagating currents of Weyl electrons of the opposite
helicities (red and blue arrows). It results in an accumulation of
right-handed (+) and left-handed (-) Weyl electrons at the two ends
of samples, respectively. }
\end{figure}

Although the introduction of the valley degree of freedom is based
upon the separation of momentum space, and looks artificial, it can
actually be as real as the other degrees of freedom such as spin,
within the certain energy and momentum scales. In particular, when
$n=1$, a pair of Weyl nodes can be mapped to the right-handed and
left-handed neutrinos respectively, albeit in a much lower energy
scale. When the system is clean enough or only has the spatially smooth
disorders, the scattering between the two valleys is negligible, and
the electron can maintain its valley identity for a sufficient long
time. In this case, one expects a valley accumulation in the boundary
of the system when there is a bulk valley current, as shown in Fig.~\ref{fig:valley-polarization}.
The detailed profile of such an accumulation depends on the condition
of the sample boundary and the valley relaxation. 

Earlier attempts to utilize the valley degree of freedom are mainly
focused on the graphene-based systems~\cite{Rycerz,gunlycke,xiao yao niu,Yao Xiao Niu2}.
More recently, the monolayer molybdenum disulphide (MoS$_{2}$) is
proposed to be an ideal material for valleytronics~\cite{DiXiaovalley2012,Cao2012}.
Our analysis shows that the Weyl metals can also be a candidate for
the valleytronics, and the valley polarization can be achieved by
applying a magnetic field.

Next, we consider the possibility for generating a real electric current
in a 3D metal by applying a strain field. A carefully engineered strain
field could induce artificial magnetic field with its magnitude and
direction depending on the momentum~\cite{Vozmediano2010,Guinea2009}.
Such an artificial magnetic field can have the opposite directions
in the different valleys, as observed in graphene~\cite{Levy2010}.
If this can be done in a Weyl metal, from Eq.~(\ref{eq:jpm}), it
will generate a real electric current.

The proper framework to consider the effect is to use the complete
form of semiclassical equation:
\begin{align}
\dot{\boldsymbol{r}} & =\bm{v}(\bm{k},\bm{r})-\overleftrightarrow{\Omega}^{\boldsymbol{kr}}\cdot\dot{\boldsymbol{r}}-\overleftrightarrow{\Omega}^{\boldsymbol{kk}}\cdot\dot{\boldsymbol{k}},\label{eq:rdot1-1}\\
\dot{\boldsymbol{k}} & =\overleftrightarrow{\Omega}^{\boldsymbol{rk}}\cdot\dot{\boldsymbol{k}}+\overleftrightarrow{\Omega}^{\boldsymbol{r}\boldsymbol{r}}\cdot\dot{\boldsymbol{r}},\label{eq:kdot1-1}
\end{align}
where $\bm{v}(\bm{k},\bm{r})=\partial\tilde{\epsilon}(\bm{k},\bm{r})/\partial(\hbar\bm{k})$,
$\Omega_{\alpha\beta}^{\bm{k}\bm{k}}=-2\mathrm{Im}\left\langle \partial u(\bm{r},\bm{k})/\partial k_{\alpha}\left|\partial u(\bm{r},\bm{k})/\partial k_{\beta}\right.\right\rangle $,
$\Omega_{\alpha\beta}^{\bm{k}\bm{r}}=-2\mathrm{Im}\left\langle \partial u(\bm{r},\bm{k})/\partial k_{\alpha}\left|\partial u(\bm{r},\bm{k})/\partial r_{\beta}\right.\right\rangle $,
and $\overleftrightarrow{\Omega}^{\bm{r}\bm{k}}$, $\overleftrightarrow{\Omega}^{\bm{r}\bm{r}}$
are defined similarly~\cite{Sundaram Niu}. Note that in the presence
of spatial inhomogeneity due to the strain field, the Bloch wave function
and the energy dispersion are in general functions of both the wave
packet central momentum $\bm{k}$ and the central position $\bm{r}$.
As a result, the new components of the Berry curvatures arise. Here,
$\overleftrightarrow{\Omega}^{\bm{r}\bm{r}}$ assumes the role of
that of the magnetic field $\bm{B}$ in Eq.~(\ref{eq:kdot1}), and
can be considered as an artificial magnetic field. 

In most of the cases the strain field is weak, and the spatial gradient
of the Bloch wave function is a small quantity. One can expand the
solution of Eq.~(\ref{eq:rdot1-1}--\ref{eq:kdot1-1}) in the orders
of the spatial gradients. To the second order of the spatial gradients,
we find that the total electric current density can be written as,
\begin{multline}
\bm{j}(\bm{r})=\bm{\nabla}_{\bm{r}}\times\bm{m}(\bm{r})-\frac{e}{\hbar}\int[\mathrm{d}\bm{k}]f(\tilde{\epsilon})\bm{\Omega}^{\bm{k}\bm{k}}\times\bm{\nabla}_{\bm{r}}\tilde{\epsilon}\\
-\frac{e}{2\pi h}\int\bm{\Phi}_{\epsilon,\bm{r}}^{(2)}f(\epsilon)\frac{\mathrm{d}\epsilon}{2\pi}\,,\label{eq:jr}
\end{multline}
where $\bm{m}(\bm{r})\equiv-(e/\hbar)\int[\mathrm{d}\bm{k}]\bm{\Omega}^{\bm{k}\bm{k}}g(\tilde{\epsilon})$
with $g(\tilde{\epsilon})\equiv\int\mathrm{d}\tilde{\epsilon}f(\tilde{\epsilon})$,
and we define a vector $\left(\bm{\Omega}^{\bm{k}\bm{k}}\right)_{\alpha}\equiv(1/2)\epsilon_{\alpha\beta\gamma}\Omega_{\beta\gamma}^{\bm{k}\bm{k}}$.
$\bm{\Phi}_{\epsilon,\bm{r}}^{(2)}$ is the second-class Chern-flux
through the momentum space iso-energy surface $S_{\epsilon,\bm{r}}$
with $\tilde{\epsilon}(\bm{k},\bm{r})=\epsilon$ at the spatial position
$\bm{r}$. Specifically, it is defined as,
\begin{equation}
\Phi_{\epsilon,\bm{r}}^{(2)\alpha}=\frac{1}{2}\epsilon_{\alpha\beta\gamma}\int_{S_{\epsilon,\bm{r}}}\mathrm{d}\bm{k}\left[\Omega_{\beta\gamma}^{\bm{r}\bm{r}}\Omega_{12}^{\bm{k}\bm{k}}+\Omega_{2\beta}^{\bm{k}\bm{r}}\Omega_{\gamma1}^{\bm{r}\bm{k}}+\Omega_{\beta1}^{\bm{r}\bm{k}}\Omega_{2\gamma}^{\bm{k}\bm{r}}\right],\label{eq:Phi2}
\end{equation}
where we have defined $k_{1}$ and $k_{2}$ as the two generalized
coordinates for parametrizing the two dimensional iso-energy surface
$S_{\epsilon,\bm{r}}$, with the normal direction of the surface defined
by $\bm{v}$. It is easy to observe that the integrand in Eq.~(\ref{eq:Phi2})
is exactly the second Chern-form of Berry curvature defined in the
manifold ($r_{1},r_{2},k_{2},k_{2}$), where $r_{1}$ and $r_{2}$
are the spatial coordinates in the plane perpendicular to the direction
of the current.

The electric current density induced by a strain field has three parts
of contributions, corresponding to each of the three terms in Eq.~(\ref{eq:jr}),
respectively: the first term is the magnetization current due to the
spatial inhomogeneity; the second term is the anomalous Hall current~\cite{Nagaosa AHE rmp}
driven by the electro-elastic potential $-\bm{\nabla}_{\bm{r}}\tilde{\epsilon}$;
and most importantly, there arises a topological contribution proportional
to $\bm{\Phi}_{\epsilon,\bm{r}}^{(2)}$. It is easy to see the total
current across a cross-section of the sample $I=\int\mathrm{d}\bm{S}\cdot\bm{j}$
has a topological contribution dictated by the second Chern number
because $\int\mathrm{d}\bm{S}\cdot\bm{\Phi}_{\epsilon,\bm{r}}^{(2)}=(2\pi)^{2}n$
with $n$ being an integer, and we have:
\begin{equation}
I_{\mathrm{topo.}}=-\frac{ne}{h}(\epsilon_{F}-\epsilon_{0}).\label{eq:Itopo}
\end{equation}

We discuss the implications of Eq.~(\ref{eq:jr}) and Eq.~(\ref{eq:Itopo})
for different kinds of strain fields. For a strain field created by
the usual lattice deformation, one expects that the artificial magnetic
field varies spatially, and there cannot be big spatial area with
nearly uniform artificial magnetic field, as it costs huge elastic
energy to sustain the required strain field~\cite{Vozmediano2010,Guinea2009}.
In this case, one cannot expect a net electric current induced by
the strain field. This does not exclude the possibility of existence
of a local current density which may in turn modify the lattice dynamics. 

A particularly interesting case is when the system has topological
defect, e.g., a disclination, one expects an artificial magnetic field
distributing around the disclination line~\cite{Vozmediano2010}.
A non-zero second Chern number in Eq.~(\ref{eq:Itopo}) implies the
existence of a chiral conducting channel along the line of disclination,
just like the chiral edge state of a quantum Hall insulator. It is
important to note that the chiral conducting channel is in a metal~,
with its robustness protected by the structure of the topological
defect and the Fermi surface topology. The construction of a model
system to realize the novel possibility of creating topologically
protected chiral modes in a metal is left for future investigations. 

In summary, we discuss the topological invariants in three dimensional
metals. We find that non-zero reciprocal space magnetic fluxes threading
through the Fermi surface will give rise to the magnetic-field-to-valley-current
response, although no electric current response is possible due to
the Fermion doubling theorem. We further observe that a strain field
could induce an electric current dictated by a second Chern number.
It implies the possibility of existence of chiral conducting channel
in a metallic system with appropriate structural topology and momentum
space topology. 

This work is supported by 973 program of China (2009CB929101, 2012CB921304).
We thank Tao Qin for useful discussion.

\end{document}